\documentclass[authoryear,12pt,3p,letter]{jowarticle}
\usepackage{graphicx}
\usepackage{amsthm,amsmath}
\usepackage{amssymb}
\usepackage{amsfonts}
\usepackage{tensor}
\usepackage{natbib}
\usepackage{setspace}
\usepackage[all]{xy}
\usepackage{enumitem}
\usepackage{titlesec}
\usepackage{mathrsfs}
\usepackage{xcolor}
\usepackage{url}
\makeatletter
\renewcommand\section{\@startsection{section}{1}{\z@}{-3.25ex plus -1ex minus -.2ex}{1.5ex plus .2ex}{\normalsize\bf}}
\renewcommand\subsection{\@startsection{subsection}{2}{\z@}{-3.25ex plus -1ex minus -.2ex}{1.5ex plus .2ex}{\normalsize\bf}}
\renewcommand\subsubsection{\@startsection{subsubsection}{3}{\z@}{-3.25ex plus -1ex minus -.2ex}{1.5ex plus .2ex}{\normalsize\bf}}
\makeatother

\newtheorem{innercustomgeneric}{\customgenericname}
\providecommand{\customgenericname}{}
\newcommand{\newcustomtheorem}[2]{%
  \newenvironment{#1}[1]
  {%
   \renewcommand\customgenericname{#2}%
   \renewcommand\theinnercustomgeneric{##1}%
   \innercustomgeneric
  }
  {\endinnercustomgeneric}
}

\newcustomtheorem{prop*}{Proposition}
\newcustomtheorem{cor*}{Corollary}

\usepackage{framed}
\usepackage{subfigure} 
\definecolor{shadecolor}{RGB}{255, 235, 200} 

\begin{document}

\begin{frontmatter}
\title{Ceci n'est pas un gluon}
\author{India Bhalla-Ladd}\ead{ibhallal@uci.edu}
\author{Eleanor March}\ead{eleanor.march@uci.edu}
\author{James Owen Weatherall}\ead{weatherj@uci.edu}
\address{Department of Logic and Philosophy of Science\\ University of California, Irvine}
\begin{abstract}
    We discuss and then resolve a tension between how physicists treat gauge bosons and the celebrated ``Wu-Yang dictionary'', which identifies particle physics terminology with that of principal bundles and principal connections.  We show how this tension leads to an interpretative choice that is not widely discussed in the physics literature.  We then show how the same considerations present a dilemma for a recent ``particle-first'' approach to Yang-Mills theory due to Henrique Gomes.  Either the particle-first approach has surplus structure as compared to principal-bundle-based approaches, or gauge bosons are not sections of vector bundles.
\end{abstract}
\end{frontmatter}
\doublespacing

\section{Introduction}\label{sec:intro}

The year 1975 saw a startling convergence of ideas developed independently in mathematics and in particle physics.  On one side were the physical theories introduced by C. N. Yang and Robert Mills in \citeyear{Yang+Mills}, which by the early 1970s had become the basis for the Standard Model of Particle Physics.  On the other side was work in the differential geometry of principal bundles and principal connections, developed by mathematicians from the late 1940s onwards. These independent strands met with the publication of the celebrated ``Wu-Yang Dictionary'' \citep{Wu+Yang}, which showed how to translate between the two argots. They showed that what physicists called a ``gauge field'' was a ``principal connection''; a ``field strength tensor'' was the ``curvature'' of a connection; and the ``gauge group'' of a theory was the structure group of a certain bundle.\footnote{\citet{Palais} and \citet{Bleecker} provide compact early presentations of Yang-Mills theory from the geometric perspective.}  The Wu-Yang dictionary sparked a fruitful collaboration between mathematicians and physicists.  In philosophy, it is the basis for the ``fiber bundle interpretation'' of Yang-Mills theory \citep{Lyre,Leeds,Catren,Healey,Arntzenius, WeatherallFBYMGR,Jacobs,GomesElement}.

Given this storied history, one might expect modern textbook treatments of particle physics to treat gluons as principal connections on an $SU(3)$ bundle.  But they do not do that.  A principal connection on an $SU(3)$ bundle is, by geometric necessity, valued in an 8-dimensional real vector space---specifically, the Lie algebra $\mathfrak{su}(3)$.  Meanwhile, gluons are usually taken to be valued in an 8-dimensional \emph{complex} vector space.  This space is also endowed with a Lie bracket, and thus forms a Lie algebra; but it is not the Lie algebra $\mathfrak{su}(3)$, rather it is $\mathfrak{sl}(3,\mathbb{C})$, which is the Lie algebra of a completely different Lie group. It is not just $\mathfrak{su}(3)$ in disguise.  

Is the Wu-Yang dictionary a myth?  As we will argue, the answer is ``not exactly''. With a little work, standard practice can be squared with Wu-Yang lore.  But there is an ambiguity that can be traced back to the Wu-Yang dictionary itself, which concerns whether we should think of a ``gauge field'' as a principal connection, or as the connection fields induced on associated bundles, relative to a choice of additional structure (gauge).  The difference is inconsequential for many practical purposes.  But for foundational purposes, the difference matters, because it is implicated in fundamental questions about what kind of entity gauge fields are, and what it means to quantize them.  As we will argue, it also bears on a very recent proposal in the philosophy of Yang-Mills theory due to Henrique \citet{Gomes}, who argues that Yang-Mills theory can be reformulated, geometrically, without using principal bundles at all.

The remainder of the paper will proceed as follows.  In section \ref{sec:fun}, we will review the geometry of Lie groups and Lie principal bundles.  Section \ref{sec:core} will elaborate and then resolve the tension discussed above. Then, in section \ref{sec:Gomes}, we introduce Gomes' ``particles-first'' approach to Yang-Mills theory  and show that the considerations we raise present a dilemma for that view. Section \ref{sec:conclusion} will conclude.

\section{Preliminaries: Lie groups, principal bundles, and all that}\label{sec:fun}

We begin by recalling some background facts.\footnote{For detaile,d see \citet{Lee}.  Assume all structures that are candidates to be smooth are smooth, and all manifolds we consider are Hausdorff and paracompact.}  A \emph{Lie group} is a group that is also a smooth manifold, in such a way that the group structure and manifold structure are compatible, i.e., for any element $h$, the map $g\mapsto g\cdot h^{-1}$ is smooth.  Associated to any Lie group is a \emph{Lie algebra}, which is a vector space endowed with a Lie bracket structure.  The Lie algebra encodes the ``infinitesimal'', or linearized, structure of the Lie group.  More precisely: since any Lie group $G$ is a group, it has an identity element, $e$; since it is a manifold, it has a tangent space at $e$, $T_eG$.  The Lie algebra $\mathfrak{g}$ associated with $G$ is modeled on this tangent space.  We endow it with a Lie bracket as follows.  For each element $g\in G$, we define a smooth map $g:G\rightarrow G$ by left action: $g:h\mapsto g\cdot h$.  For each vector $\xi^a \in T_eG$, we define a vector field on $G$ defined by assigning, to each point $g$, the vector $g_*(\xi^a)$.  We thus get a family of vector fields on $G$.  One can show that these fields are closed under Lie derivatives, i.e., for any such vector field $\xi^a$ and $\eta^a$, $\mathcal{L}_{\xi}\eta^a$ is also a in the family.  We use these Lie derivatives to induce a Lie bracket on $T_eG$. Thus we find that, given a Lie group, the Lie algebra is entirely and uniquely determined.

The construction just described is completely general.  We will be interested in a particular class of Lie groups known as \emph{matrix groups}, which are typically defined by the behavior of invertible linear transformations acting on a given finite-dimensional vector space.  Take, for instance, the \emph{general linear groups}, which consist of all invertible linear transformations from a vector space $V$ to itself.  Up to isomorphisms, these groups are determined by the dimension of $V$ and the field over which $V$ is defined, so for instance, $GL(n,\mathbb{R})$ (respectively, $GL(n,\mathbb{C})$) would be the group of automorphisms on an $n$ dimensional real (complex) vector space, which form a real (complex) manifold of dimension $n^2$.  Its Lie algebra $\mathfrak{gl}(n,\mathbb{R})$ (respectively, $\mathfrak{gl}(n,\mathbb{C})$) is the $n^2$ dimensional real (complex) vector space consisting of all $n\times n$ matrices, with Lie bracket given by the commutator.  The \emph{special linear groups} $SL(n,\mathbb{R})$ and $SL(n,\mathbb{C})$, are subgroups of these consisting of matrices of determinant 1.  These are, respectively, real and complex manifolds of dimension $n^2-1$; their associated Lie algebras, $\mathfrak{sl}(n,\mathbb{R})$ and $\mathfrak{sl}(m,\mathbb{C})$ are real and complex vector spaces, respectively, of dimension $n^2-1$.

Of particular interest for particle physics are the groups $SU(n)$, determined by the unitary matrices of determinant 1 acting on an $n$-dimensional complex vector space.  These form $n^2-1$ dimensional \emph{real} manifolds---even though they are matrices acting on complex vector spaces.  How does this happen?  The reason is that the defining property of a unitary transformation $U$ is that $U^*=U^{-1}$, i.e, that its conjugate transpose is also its inverse.  As one can readily show, this condition strongly constrains the relations between the real and imaginary parts of the matrix components.  The associated Lie algrebras $\mathfrak{su}(n)$ are $n^2-1$ real vector spaces.  

A \emph{representation} of a group $G$ on a vector space $V$ is a group homomorphism $\varphi:G\rightarrow GL(V)$.  A representation is called \emph{faithful} if the homomorphism is injective.  In this case, the image of $G$ under $\varphi$ can be thought of as a collection of matrices whose algebraic structure, under ordinary matrix multiplication, is precisely that of the group $G$.  Several types of faithful representation are of special interest.  The first are the so-called \emph{fundamental} representations of matrix groups, which are those representations that map an abstract group into the matrices used to define the group.  So, for instance, the fundamental representation of $SU(n)$ would be the group homomorphism $\varphi:SU(n)\rightarrow GL(n,\mathbb{C})$ taking each element of $SU(n)$ to a special unitary matrix acting on some $n$ dimensional complex vector space.  Similarly, we have \emph{anti-fundamental} representations for matrix groups, which are corresponding representations on the (conjugate) dual space to the fundamental representation.  The fundamental and anti-fundamental representations extend (factorwise) to representations on arbitrary tensor products of those the vector spaces.\footnote{See \citet{Coleman} for a famous and detailed discussion of representations of $SU(3)$ constructed from tensor products.}

We also have the \emph{adjoint representation} of a Lie group on its own Lie algebra, understood as a vector space.  To define it, we first define a family of smooth maps $\Upsilon_g:G\rightarrow G$, whose action on group elements is $\Upsilon_g:h\mapsto ghg^{1}$.  Then the adjoint representation of $G$ on $\mathfrak{g}$ is defined, for each $g\in G$, by $g\mapsto ((\Upsilon_g)_*)_{|e}$, i.e., the invertible linear transformation at the identity of $G$ determined $\Upsilon_g$. This representation is faithful for all groups considered in the present paper.  As with Lie algebras, there is no freedom in defining adjoint representations.  They are wholly determined by the Lie group structure.

Finally, we turn to bundles.  Recall that a \emph{smooth fiber bundle} is a smooth surjective map $\pi:B\rightarrow M$ with the property that each point $p\in M$ has a neighborhood $U$ for which $\pi^{-1}[U]$ is diffeomorphic to $U\times F$, for some fixed manifold $F$, in such a way that $\pi$ coincides, under the action of this diffeomorphism, with projection onto the first factor.  Here $B$ is called the \emph{total space}, $M$ is the \emph{base space}, $F$ is the \emph{typical fiber}, and $\pi$ is the projection.  Local diffeomorphisms to product manifolds of the form $U\times F$ are known as \emph{local trivializations}.  A (local) \emph{section} of a fiber bundle is a smooth map $\varphi:U\rightarrow B$, where $U$ is some subset of $M$, such that $\pi\circ \varphi = 1_M$, the identity on $M$.  The preimage of a point $p\in M$ under $\pi$ is the \emph{fiber at $p$}; the kernel of the pushforward along the projection map at a point $x\in B$ is known as the \emph{vertical space} at $x$, denoted $V_xB\subseteq T_xB$.  The vertical space at $x$ consists of those vectors at $x$ tangent to the fiber at $\pi(x)$.  

A \emph{principal $G-$bundle}, for $G$ a Lie group, is a smooth fiber bundle $P\rightarrow M$ carrying a smooth, fiber-preserving right $G$ action on $P$, which is such that for any points $p\in M$ and $x,y\in\pi^{-1}[p]$, there exists a unique $g\in G$ such that $y=x\cdot g$.  It follows from this definition that the typical fibers of a principal bundle are $G$-torsors, i.e., manifolds diffeomorphic to $G$, but carrying no group structure. The vertical space at each point of a principal bundle has the same dimension as the Lie algebra associated with $G$; in fact, the $G$ action on $P$ determines a canonical map from the vertical space at each point to the Lie algebra $\mathfrak{g}$.  In this way, the vertical spaces can be seen as copies of the Lie algebra, with induced Lie bracket given by this map, each carrying an adjoint representation of $G$.  

\section{What is a Gluon?}\label{sec:core}

We now turn to the central concern of this paper, which is to identify and then resolve an apparent tension between the Wu-Yang dictionary and standard discussions of gauge bosons in particle physics.  We will focus on the gluon, the gauge boson associated with quantum chromodynamics, though what we say could be applied, \emph{mutatis mutandis}, to W and Z bosons or photons.   Note that while we say we are focusing on \emph{quantum} chromodynamics, all of our remarks will be fully classical---as is the relevant discussion in \citet{Wu+Yang}. (We return to this issue in section \ref{sec:res}.)

\subsection{The textbook account}\label{sec:text}

We will begin by presenting a generic ``textbook'' account of gluons.\footnote{Compare with \citet{Peskin+Schroeder}, \citet{Zee}, \citet{Schwartz}, or \cite{Tong}.}  While physicists often point to the Wu-Yang dictionary, and use it when need for, e.g., non-trivial spacetime topologies, standard introductions to Yang-Mills theory do not fuss with specifying bundles or the spaces in which fields are valued.  Instead, one starts by defining the theory's dynamics by writing down a Lagrangian.  So we will begin there, and then work backwards. 

Consider the Yang-Mills Lagrangian for an $SU(3)$ gauge theory coupled to a spinor field -- say, quarks in QCD -- written in relatively standard notation:\footnote{Compare with \citet[Ch. 2]{Tong} and \citet[Ch. 25]{Schwartz}.}
\begin{equation}
    \mathcal{L}_{YM} = -\frac{1}{4}\sum_a (\partial_\mu A^a_\nu - \partial_\nu A^a_\mu + g f^{abc}A^b_\mu A^c_\nu)^2 + \Bar{\psi}(i \gamma^\mu (\partial_\mu - ig A^a_\mu T^a) )\psi 
\end{equation}
Here $\partial_{\mu}$ is a partial derivative operator (to be discussed more below), $\gamma^{\mu}$ refers to the Dirac $\gamma$ matrices, $\psi$ is a spinor field, and $g$ is a (real) coupling constant.  Following particle physics conventions, the Greek indices are spacetime indices, and the power of two is a Minkowski spacetime inner product.  The objects $T^a$ are a fixed basis for the Lie algebra associated with the theory -- in this case is $\mathfrak{su}(3)$ -- with Latin indices $a,b,c = 1, 2, \ldots, 8$ labeling those basis elements.  (For these objects, no distinction is drawn between raised and lowered indices.) The object $f^{abc}$, which is totally antisymmetric under index permutations, represents the ``structure constants'' for the Lie algebra, which encode its algebraic structure as follows: $i f^{abc} T^c = [T^a, T^b]$.  These basis elements are often called ``generators'' for the algebra.  Repeated indices in any term are implicitly summed.

This leaves the fields $A^a_{\mu}$.  There are eight such fields, each a smooth spacetime covector field (i.e., varying with spacetime location), labeled by the counting index $a$.  It is these fields together that are often called ``gluons''.\footnote{Compare with: ``We will refer to the fields $A^a_{\mu}$ collectively as \emph{gluons}, in deference to the fact that the strong nuclear force is described by G = SU(3) Yang-Mills theory'' \citep[p. 28]{Tong}.  Or consider \citet[pp. 491 \& 508-9]{Schwartz}, who refers to these as the ``gauge fields'', though he mentions elsewhere that they determine a connection (p. 489).}  The individual gluon fields are dependent on the basis $T^a$, though the sum $A^a_\mu T^a$ is a well-defined Lie-algebra-valued one-form.  In addition, the fields are gauge-dependent in the following sense.  The Lagrangian $\mathcal{L}_{YM}$ is invariant under local (infinitesimal) transformations of the form: 
\begin{subequations}\label{gaugeTrans}
\begin{eqnarray}
    \psi(x) &\mapsto& e^{i\alpha^a (x) T^a}\psi(x) \\
    A^a_\mu(x) &\mapsto& A^a_\mu(x) + \frac{1}{g}\partial_\mu \alpha^a(x) - f^{abc}\alpha^b(x) A^c_\mu (x)
\end{eqnarray}
\end{subequations}
where $\alpha^a(x)$ are components of a Lie algebra element expressed in the basis $T^a$.  These are (infinitesimal) \emph{gauge transformations}.  

What kind of objects are the gluons, then?  Evidently, they are components of Lie-algebra-valued one-forms on spacetime.  They are defined using basis elements of $\mathfrak{su}(3)$.  And so one might infer that they live in, or define one-forms valued-in, the vector space $\mathfrak{su}(3)$.  But this cannot be quite right, given what has been said already.  This is because these are evidently the sorts of objects that can be multiplied by imaginary numbers, and whose commutators involve imaginary contributions, as reflected in the definition of the structure constants above.  And so, it would seem, these fields determine elements of an eight dimensional \emph{complex vector space}.  But as we saw in the previous section, $\mathfrak{su}(3)$ is, by mathematical necessity, an eight dimensional \emph{real} vector space.  

In fact, as we said in section \ref{sec:intro}, the gluons are components of a Lie-algebra-valued one-form valued in the \emph{complexification} of $\mathfrak{su}(3)$, which is the complex vector space freely generated by multiplying basis elements of $\mathfrak{su}(3)$ by complex scalars.  The Lie algebra structure on $\mathfrak{su}(3)$ lifts to a Lie algebra structure on this vector space, too, which is actually the algebraic structure determined by the structure constants.  This Lie algebra is $\mathfrak{sl}(3,\mathbb{C})$.\footnote{We are hardly the first to notice this. See, for instance, \citet[p. 507]{baez2010algebragrandunifiedtheories}.}

\subsection{The Wu-Yang Fictionary?}

The conclusion of the previous subsection should give one pause, for two reasons.  The first is that it is nearly ubiquitously claimed that gluon fields are valued in, or transform under, the adjoint representation of $SU(3)$.  Here is a typical statement:
\begin{quote}
    The adjoint representation is of particular importance in particle physics. In Yang-Mills theory, the gauge bosons, the analogs of the familiar photon, belong to the adjoint representation. \citep[p. 241]{Zee}
\end{quote}
But as we have seen, the adjoint representation of $SU(3)$ is necessarily a real vector space, whereas the fields identified with gluons are complex.  

One might write this off to harmless sloppiness.  As noted, gluons are valued in the \emph{complexified} adjoint representation, i.e., another representation of $SU(3)$ generated from the adjoint in a straightforward way.  In that sense, the adjoint action fully determines the transformation properties of gluon fields.  But there is a deeper issue here, which is that there are independent reasons to think that gluons should belong to the non-complexified adjoint representation.  These reasons have to do with the geometrical interpretation of Yang-Mills theory as described in the Wu-Yang dictionary.  According to that interpretation, gauge fields should be identified with a \emph{principal connection} on a certain principal bundle.

A \emph{connection} on a fiber bundle $\pi:B\rightarrow M$ is additional structure, not determined by the bundle structure, that provides a local ``standard of constancy'' for the purposes of differentiation.  More precisely, a \emph{connection} is a smooth, vertical-valued one-form $\omega^{\alpha'}{}_{\beta}$ on $B$, which acts as a projection on the vertical space in the sense that $\omega^{\alpha'}{}_{\beta}\omega^{\beta'}{}_{\kappa}=\omega^{\alpha'}{}_{\kappa}$ and $\omega^{\alpha'}{}_{\beta}[T_xB]=V_xB$.\footnote{The notation here follows \citet{Geroch} and \citet{WeatherallFBYMGR}; note the change from the previous subsection.  Briefly, lower case Latin indices are used for tensors associated with the base space of a bundle (e.g., spacetime); Greek indices are used for tensors valued in the tangent spaces of the total space; and a ``prime'' symbol indicates that an index is vertical.  We will presently use mathfrak indices for Lie algebra elements and upper case Latin indices for vectors in vector bundles associated to a principal bundle.}  A connection determines a smooth distribution of horizontal (i.e., everywhere non-vertical) subspaces of the tangent spaces at points of $B$, consisting of the kernel of $\omega^{\alpha'}{}_{\beta}$ at each point.  In the special case of a principal $G$-bundle $\pi:P\rightarrow M$, we define a \emph{principal connection} to be a connection that is compatible with the principal bundle structure.  First, as we have already observed, the vertical space at each point of a principal bundle is canonically isomorphic to the Lie algebra of $G$, which means that any connection $\omega^{\alpha'}{}_{\beta}$ on $P$ can be understood as a Lie-algebra-valued one-form, $\omega^{\mathfrak{A}}{}_{\beta}$. This field is called a principal connection if, for any $g\in G$, $(R_{g^{-1}})^*(\omega^{\mathfrak{A}}{}_{\alpha})=g^{\mathfrak{A}}{}_{\mathfrak{B}}\omega^{\mathfrak{B}}{}_{\alpha}$.  Here $R_{g^{-1}}$ is the smooth action of $g^{-1}$ on $P$, and $g^{\mathfrak{A}}{}_{\mathfrak{B}}$ is the adjoint action of $g$ on the Lie algbra $\mathfrak{g}$.  In other words, a principal connection is a connection whose variation along the fibers is fully determined by the group structure acting on the fibers.

It is important that a principal connection is valued in the adjoint representation of $G$, not the complexified adjoint representation (for $G$ real).  This is because the complexified adjoint is not isomorphic to the vertical spaces at each point of $P$.  A field valued in the complexified adjoint is not a principal connection, because it is not a connection at all.  It follows that, insofar as a gluon is a Lie-algebra-valued one-form valued in the complexified adjoint representation of $SU(3)$ on $\mathfrak{sl}(3,\mathbb{C})$, it is not a principal connection on an $SU(3)$ bundle.  This is the tension between textbook accounts and the Wu-Yang dictionary that we identified in the introduction.  In the next subsection, we will resolve the tension---though in doing so, we will face an interpretive choice that has significance for other foundational issues.

\subsection{The resolution, and an interpretive choice}\label{sec:res}

Having now made a big deal of this tension, we freely acknowledge that its resolution is straightforward---though we maintain that it raises important issues.  We need just a little more bundle-ology. 

Fix a principal $G-$bundle $\pi:P\rightarrow M$ and consider a vector space $V$ carrying a representation $\varphi$ of $G$.\footnote{It is generally convenient to assume that $V$ carries a preferred basis.  See \citet{Kobayashi+Nomizu} and \citet[Appendix A]{WeatherallFBYMGR}.}  We define a new bundle, with typical fiber $V$, by taking $P\times V$ and quotienting by the following $G$ action: $(x,v)\mapsto (xg,\varphi(g^{-1})v)$.  The resulting bundle, $P\times_G V\rightarrow M$, is called an \emph{associated vector bundle}.  If $G$ is a matrix group acting on $V$, then this construction induces additional structure, preserved by $G$, on the fibers of $P\times_G V$.  To take an example, if we begin with an $SU(3)$ bundle and consider the fundamental representation of $SU(3)$ on $\mathbb{C}^3$, the corresponding associated vector bundle has as fibers three dimensional complex vector spaces \emph{endowed with a Hermitian inner product and preferred orientation}.  Sections of associated bundles represent matter fields subject to Yang-Mills forces, such as quarks.  Different fields participating in a force may be valued in different bundles carrying different representations of $G$, all associated to one principal bundle.  

Matter dynamics are given by differential equations imposed on sections of associated bundles. It is a fact of life on bundles that derivatives of sections can only be taken relative to some choice of derivative operator, of which there are generally many.  This is where principal connections come in. Any principal connection on $P$ induces a unique covariant derivative operator $\nabla$ on sections of associated bundles, and vice versa.   Given any two derivative operators $\nabla$ and $\nabla'$ acting on sections of a vector bundle, there exists a unique field $C^A{}_{Bc}$ on $M$ with the property that for any section $\psi^A$, $\nabla_a \psi^A = \nabla'_a\psi^A - C^A{}_{Ba}\psi^B$.  This expression says that the difference in action between any two derivative operators, in any direction on $M$, is given by an \emph{endormorphism} field, i.e., a linear transformation from each fiber to itself. The field $C^A{}_{Bc}$ is thus called an endomorphism-valued form.

With these observations in mind, consider the Yang-Mills Lagrangian, $\mathcal{L}_{YM}$.  This Lagrangian is best understood as a (second-order) Lagrangian on a complex spinor bundle whose sections are the fields $\psi$.  The gauge transformations, Eqs. \eqref{gaugeTrans}, involve an $SU(3)$ action on the fibers of the bundles, which preserves the Hermitian inner product used to take scalar products of the fields.  Derivatives are, by necessity, taken relative to a derivative operator. But what derivative operator?  In fact, there are \emph{two} derivative operators that appear in the Lagrangian.  One is written $\partial_\mu$.  This is a flat derivative operator, generally determined by a local trivialization.  The other is $\partial_{\mu} - ig A^a_\mu T^a$.\footnote{In this expression we return to the particle physics notation from section \ref{sec:text}.}  The way to understand this latter expression is as defining a covariant derivative operator, using the fixed derivative $\partial_{\mu}$ and an endomorphism-valued form $ig A^a_\mu T^a$, as above.  Why does this field have a different index structure than $C^A{}_{Bc}$?  In fact, it does not; rather, some indices have been suppressed.  The important fact is that $\mathfrak{sl}(3,\mathbb{C})$ happens to consist of all trace-free complex 3x3 matrices, each of which is an endormorphism acting on the bundle in which $\psi$ is valued.

Given this, how should we think about the gluon fields $A^a_\mu T^a$?  They are not a principal connection.  Instead, they sum (over $a$) to give the endomorphism-valued form relating a derivative operator to a fixed background flat derivative operator, $\partial_{\mu}$.  In that sense, they define, or determine, a principal connection.  But they do so only relative to a choice of background derivative operator.  Changing that background derivative operator can be thought of as implementing a gauge transformation---which in turn requires a change both in the representation of $\psi$ and of $A^a_{\mu}T^a$.  The fields $A^a_{\mu}$ are components of that endomorphism-valued form in the basis given by the generators $T^a$.  Confusingly, both $A^a_{\mu}$ and $A^a_{\mu}T^a$ are also sometimes called connections or, in the former case, ``connection coefficients''.  But they are importantly different from a connection as we have defined it and, most importantly, they are emphatically \emph{not} a principal connection on a principal bundle.  They give rise to one only relative to the choice of $\partial_{\mu}$.

\begin{figure}[h!]
    \centering
    \includegraphics[width=0.75\linewidth]{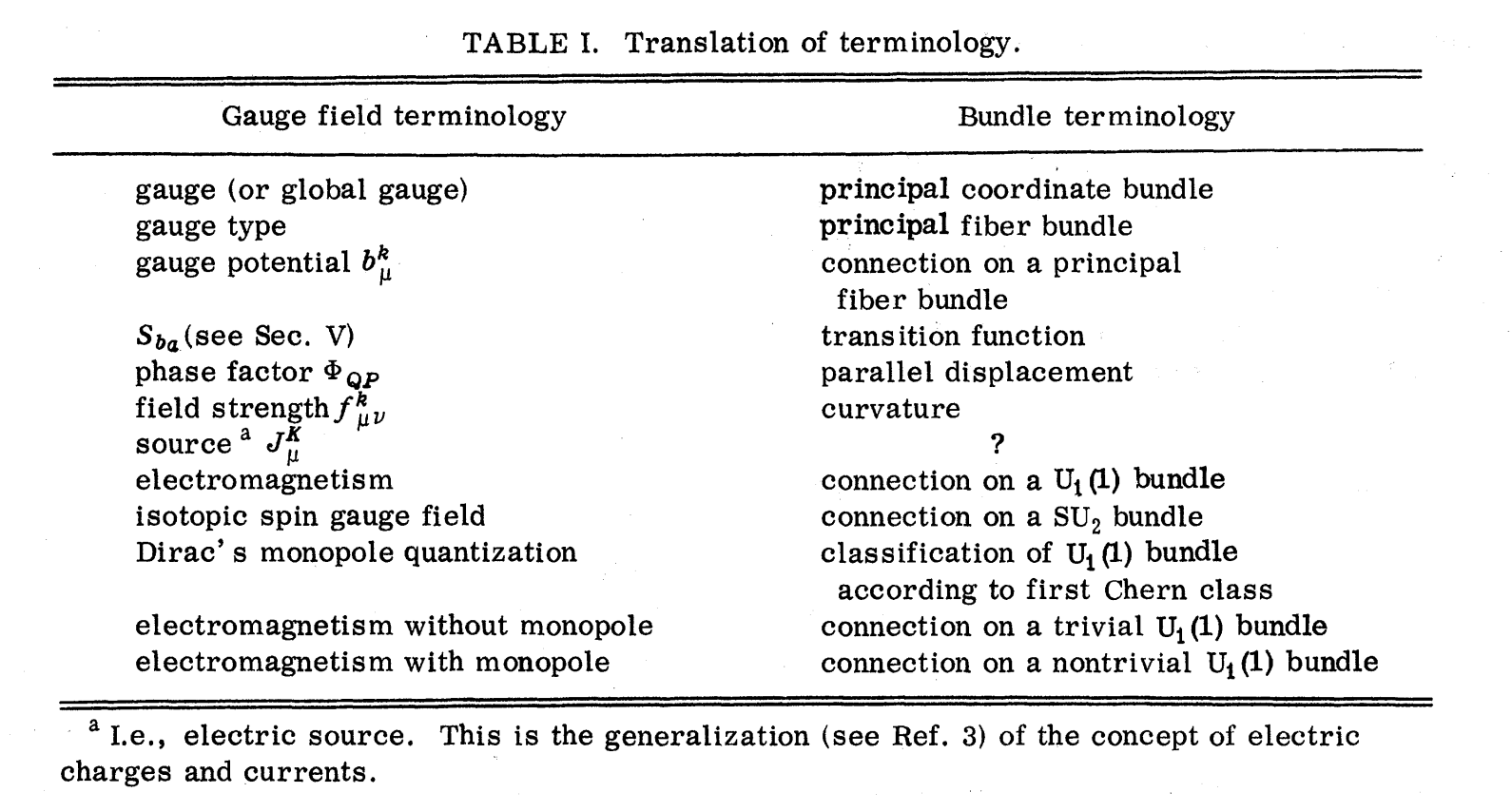}
    \caption{Translation table from the original Wu-Yang dictionary.}
    \label{fig:placeholder}
\end{figure}

In fact, we can see the ambiguity between gauge-fields-as-principal-connections and gauge-fields-as-connection-coefficients already in the original Wu-Yang paper.  As shown in Fig \ref{fig:placeholder}, Wu and Yang identify the ``gauge potential'', i.e., the field $A^a_{\mu}T^a$, with the ``principal connection on a principal fiber bundle''.  But this is not a strict translation.  As we have seen, $A^a_{\mu}T^a$ encodes information about the principal connection, but only in the presence of a fixed background trivialization or flat derivative operator.   

Now that we are clear on how these objects relate to one another, we come to the interpretive choice noted above.  Should we treat gluons as complex gauge fields $A^a_\mu T^a$, or as a connection on a principal bundle determined by those fields?  Although physicists often do identify gluons just with the local gauge fields, doing so is problematic for a number of reasons, chief among which is that the fields representing a physical entity would be gauge-dependent.  The alternative option, which is the key move in the geometric interpretation of Yang-Mills, is to take those fields to define a gauge \emph{invariant} object, viz., the covariant derivative operator.  The degrees of freedom of this object, relative to a fixed choice of $\partial_{\mu}$, coincide with the local gauge fields, but the mathematical object that represents some aspect of physical reality on this view would be the derivative operator as a whole---or, perhaps better, the principal connection that induces it.

This choice has important consequences for the interpretation of the theory.  For instance, as \citet{Trautman} and \citet{WeatherallFBYMGR} point out, taking the derivative operator to represent the entity we otherwise think of as the ``gluon field'' leads to strong analogies between the geometry and physical interpretation of Yang-Mills theory and that of general relativity.  But on the other hand, it also presents challenges.  In standard approaches to quantizing Yang-Mills theory, the fields $A^a_{\mu}T^a$ are treated as the classical fields whose quantum analogues one wishes to study.  It is, therefore, the ``quanta'' of the gauge field that are the particles of the gluon field.  Indeed, there seems to be no widely accepted notion of a ``quantum derivative operator'' or ``quantum principal connection''.  In our view, developing such a notion should be a high priority for foundationally-oriented physicists.  In the meantime, there is a plausible argument that the Wu-Yang dictionary and geometric interpretation of Yang-Mills theory do not fully carry over to the quantum case.

\section{Gomesification}\label{sec:Gomes}

We now show how the remarks in the forgoing sections bear on a recent proposal due to Henrique \citet{Gomes}.  On Gomes's ``particle-first'' approach to Yang-Mills theory, one forgoes talk of principal bundles altogether.  Instead, one works directly with vector bundles.  The role that principal bundles play on more traditional approaches, of relating different associated bundles, is replaced by using fibered tensor products of vector bundles carrying the fundamental and anti-fundamental representations of the theory's structure group.  So, for instance, for an $SU(3)$ theory, one begins with a base manifold $M$ and considers a vector bundle $E\rightarrow M$ whose fibers are three-dimensional complex vector spaces $V$ carrying a Hermitian inner product and preferred orientation, and then one considers various bundles whose fibers are tensor products of that space and its conjugate dual.  One uses the fiber-wise tensor product construction instead of the associated bundle construction to show how these bundles are all related to one another.  Thus, one need only encounter bundles whose sections represent matter fields, rather than principal bundles, whose representational significance is more abstract.

Gomes highlights that on his view, the gauge bosons become ``...of a piece with other matter fields in that they are all sections of (different) tensor bundles over the same underpinning vector bundle'' \citep[p. 8]{Gomes}.  Specifically, they are sections of the bundle of trace-free endomorphism-valued forms on $M$.  This means gluons are identified with $\mathfrak{sl}(3,\mathbb{C})-$valued forms, just as we saw in the textbook account above.  Doing so carries all of the costs we have already noted.  Most importantly, for Gomes's project, it requires the specification of a background flat derivative operator, relative to which his gluon fields can determine a covariant derivative acting on other fields, such as quarks.  In this sense, then, his representation of gluons requires additional structure, beyond just the vector bundles and dynamical sections thereof, since his gluons fields can play their full representational role only relative to an additional choice.  This choice is conventionally given by a preferred local trivialization of the fundamental vector bundle, i.e., a choice of gauge.  Thus, the Gomesified representation of gauge bosons is gauge dependent.  

In later work, Gomes acknowledges that his representation of gauge bosons depends on a preferred trivialization.
\begin{quote}
   Thus, given a trivialisation ... the connections parametrise the space of covariant derivatives. It is only at this step -- after a local trivialisation -- that the covariant derivatives are described as vector bosons: 1-forms valued on $End(E)$.... \citep[p. 15]{GomesAB}
\end{quote}
He emphasizes that this can only be done locally, and acknowledges that that is a shortcoming.  But we think there is a deeper worry, related to the interpretive choice we discussed at the end of the previous section.  On the one hand, he could reify the endomorphism-valued forms as representing the real physical degrees of freedom of a gauge boson, at the cost of introducing something like a preferred gauge.  This option makes gauge bosons vector fields, but at the cost of adding excess structure. The other option is to take the covariant derivative operator on the fundamental vector bundle to represent gauge bosons. This option would be closer to saying that it is the principal connection, as opposed to the connection coefficients, that represents gauge bosons.  But it comes at a cost, insofar as gauge bosons would no longer be sections of vector bundles.\footnote{In personal correspondence, and in some talks, Gomes has made clear that he currently prefers the second option.}

Once we allow for derivative operators on those bundles to represent physically significant structure, it becomes harder to hold on to the differences are between a principal bundle approach and a vector bundle approach, where one understands the former along the lines of, say, \citet{WeatherallFBYMGR}, who takes the principal bundle to be nothing more than a convenient way of encoding the derivative operators shared between different associated bundles.  If all that is at stake in the debate is whether a certain covariant derivative operator should be lifted from one vector bundle to other, systematically related ones by means of a principal bundle to which they are all associated or by means of a tensor bundle construction implementing precisely the same relationships, then it seems best to acknowledge that both options are available, that they carry the same physical information about the world and have the same significance, and leave the rest to taste.

\section{Conclusion}\label{sec:conclusion}

We have highlighted an apparent tension between standard practice in textbook accounts of Yang-Mills theory and the geometric formulation of the theory as characterized in the Wu-Yang dictionary, and we have shown how this tension can be resolved by distinguishing between principal connections as Lie-algebra valued forms on a principal bundle and the possibly-complex-valued connection coefficients determining (complex) endomorphism-valued forms acting on associated bundles.  These are related to principal connections and to the adjoint representation of the associated structure group, but they are not themselves principal connections and they are not valued in the adjoint representation.  Moreover, to get from one to the other requires one to fix additional structure, viz., a background flat derivative operator.

\section*{Acknowledgments} This material is based upon work supported by the National Science Foundation under Grant No. 2419967. We are grateful to Michael Ratz and Henrique Gomes for discussions about this material, and to Gomes for comments on an earlier draft.

\bibliographystyle{elsarticle-harv}
\bibliography{refs}

\end{document}